# Multispeckle diffusing wave spectroscopy as a tool to study heterogeneous mechanical behavior in soft solids

Jianzhu Ju [a], Luca Cipelletti [b,c], Stephan Zoellner [d], Tetsuharu Narita*,[a,e], Costantino Creton*,[a,e]

[a] Sciences et Ingénierie de la Matière Molle, CNRS UMR 7615, ESPCI Paris, PSL Université, Paris, France

[b] Laboratoire Charles Coulomb (L2C), University of Montpellier, CNRS, Montpellier, France

[c] Institut Universitaire de France

[d] TESA SE, 22848 Norderstedt, Germany

[e] Global Station for Soft Matter, Global Institution for Collaborative Research and Education, Hokkaido University, Sapporo, Japan

## Abstract

Multiple speckle diffusing wave spectroscopy (MSDWS) can be applied to measure spatially heterogeneous mechanical behavior in soft solids, with high sensitivity to deformation and both spatial and temporal resolution. In this paper, we discuss the mathematical approach behind the quantification of the deformation rate from MSDWS data and provide guidelines for optimizing the selection of experimental parameters in measurements. After validating the method in extensional tests on an elastomer, we provide an example of the potentiality of MSDWS by measuring the spatial distribution of the deformation rate during shear debonding of adhesive tapes. We quantitatively characterize the deformation rate distribution related to shearing and peeling under loading. A highly heterogeneous deformation rate distribution is observed, and time-dependent measurements reveal an increase in deformation localization hundreds of seconds before full debonding. This behavior, previously predicted by theory and simulation, is here demonstrated experimentally for the first time.

*correspondence authors: tetsuharu.narita@espci.fr, costantino.creton@espci.fr

# I. Introduction

The mechanical response of soft materials under external loading provides valuable information about the structural composition and the application properties under dynamic conditions, especially for soft materials with high deformability and viscoelasticity. In common tests on soft solids, such as shear rheology and uniaxial stretching, one assumes that the material properties are spatially homogeneous, so that the macroscopic stress under deformation (or macroscopic strain under creep stress) completely defines the mechanical response [1, 2]. However, it is commonly reported, especially at large deformations [3, 4] or in the presence of flaws [5-7], that the material properties and in particular their mechanical response can be spatially heterogeneous and deviate from that inferred from simple models assuming homogeneity.

Although measurements of the macroscopic mechanical response can also unveil local failure [8-10] and structural transitions [11, 12], the most direct and reliable way to investigate such phenomena is the full-field measurement on the whole sample, with both temporal and spatial resolution. Several imaging methods such as digital image correlation (DIC) [13], birefringence imaging [14, 15], and mechanophore fluorescence mapping [16, 17] can provide spatially-resolved information on the mechanical response. Among these methods, DIC measures the spatially-resolved deformation field and is commonly used for mechanical measurement in soft materials, since it can be easily combined with ordinary testing systems [18, 19]. DIC quantifies the displacement of region of interest (ROIs) of the sample by comparing two images of the sample surface. To optimize the DIC algorithm, the surface should be spray-painted with a highly contrasted, speckled pattern. However, the observations of spray-painted samples bring several limitations: (1) DIC is essentially a surface measurement, where the bulk deformation cannot be detected. In fact, any contribution due to the bulk needs to be minimized, to avoid interfering with the surface imaging. Furthermore, (2) spray painting may be difficult or even impossible for some materials, such as hydrogels with a high water content[5]. Sprayed speckles can also lose their original shape at large strains and detach from the sample surface, making the correlation of images difficult[20]. Finally, (3) although DIC provides in general excellent spatial resolution and high sensitivity to displacement, both spatial resolution and sensitivity are limited by the characteristic size of the speckled pattern, which can be difficult to reduce, e.g. as needed when working with small sample surfaces.

Multiple speckle diffusing wave spectroscopy (MSDWS) uses a similar way to measure mesoscopic deformation. In contrast to usual DIC, however, no spray painting is needed: speckles are generated from multiply scattered light upon illuminating a turbid sample with coherent laser light [20-22]. The intensity correlation of speckled patterns is extremely sensitive to the displacement of the scatterers, down to the scale of nanometers [23, 24]. Accordingly, MSDWS has been applied to dynamics characterization in soft [22, 25], granular [26, 27] and polymeric [20, 28] materials, taking advantage of its excellent time resolution and displacement sensitivity. In this work, we apply MSDWS to quantitatively measure the spatially heterogeneous deformation rate of soft solids. The main purpose of this work is to explore under what conditions MSDWS may be applied and explain the requirements and limitations related to the method, making it more accessible to other users.

The rest of the paper is organized as follows. In section II, we recall the general theoretical background of MSDWS and its application to the measurement of heterogeneous deformation. Focusing on the case of soft solids, we show how to simplify the mathematical processing to directly obtain spatially revolved strain rate from in-situ imaging, addressing practical issues commonly associated to this kind of measurements. In section III, we introduce the set-up and the preparation of the material. In section IV, we describe uniaxial extension tests on an elastomer that demonstrate and validate the method. Finally, in section V, we apply the method described in this work to obtain spatial maps of the deformation rate during debonding of an adhesive tape, a quantity that was experimentally inaccessible in previous works on adhesion.

## II. Measurement of heterogeneous mechanical behavior by MSDWS

## A. Testing geometry

In this work, we focus on MSDWS in the backscattering geometry [23, 24]. The experimental set-up is schematically shown in Fig. 1(a). An expanded laser beam homogeneously illuminates the surface of a sample which contains a sufficient amount of scatterers. The backscattered light is collected by a high-speed camera, which makes an image of the sample surface. Due to the coherence of the laser light, the images have a speckled appearance (inset of Fig. 1(a)). The speckles result from the interference between photons that have penetrated in the sample and have been scattered due to local fluctuations of the material refractive index. These fluctuations may arise from concentration or composition fluctuations of the raw material, or may be due to the addition of probe particles with a refractive index different from that of the material. For the turbid samples that we shall consider here, the majority of the backscattered light is due to multiple scattering. Multiple scattering is characterized by the photon transport mean free path $l^*$, the length scale over which a photon is scattered a sufficient number of times for its propagation direction to be randomized [6, 24]. $l^*$ typically ranges from a few μm up to hundreds of μm. Although the camera images the sample surface, the speckle pattern depend on the path of the photons in the bulk of the sample, such that the method is sensitive to deformations occurring in a sample layer of thickness up to about $10l^*$ [29, 30]. This makes the method suitable for mapping the in-plane deformation of a 3D slab of the material.

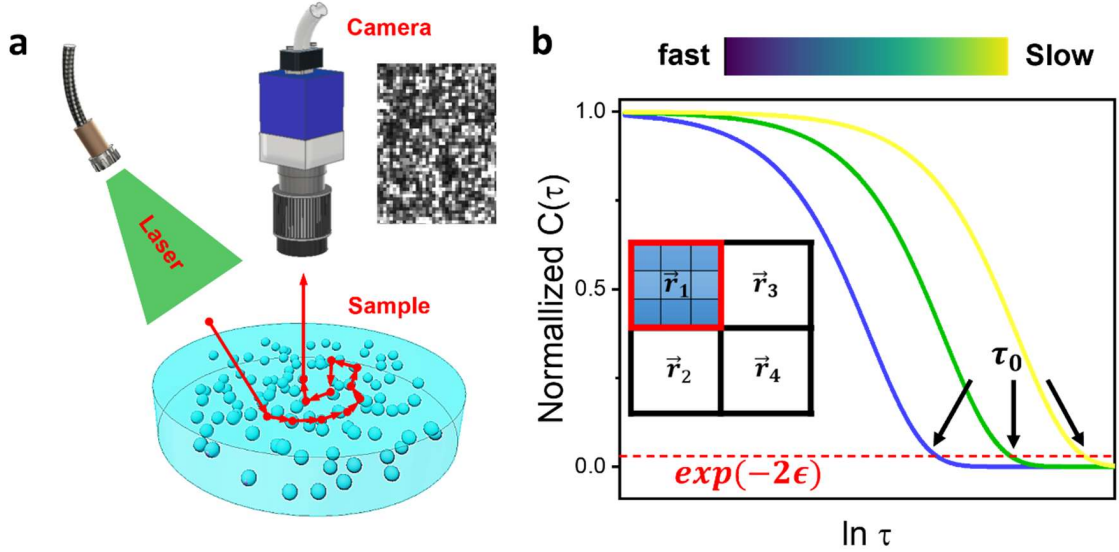

*Fig. 1(a) Scheme of a MSDWS experiment in the backscattering geometry. (b) Autocorrelation functions for systems with different dynamics. The value $exp\,(-2\epsilon)$ is indicated by the red dashed line: the intersection of the autocorrelation functions with this line defines the value of $\tau_0$. Insert: schematic view of the regions of interest (ROIs) for data processing. In the scheme, one ROI contains 3×3 pixels, centered at $\vec{r}_i$.*

## B. Autocorrelation function

### 1. Autocorrelation function in multispeckle configurations

When the scatterers are in motion due to thermal energy or owing to an external drive, the intensity of the scattered light fluctuates with time, at a rate related to the rate at which scatterers are displaced. The dynamics of the scatterers can be thus extracted by quantifying the rate of intensity fluctuations, which is achieved by calculating the autocorrelation function of the backscattered light intensity measured by the camera [21, 23, 24, 31, 32]:

$$C(\vec{r},t,\tau) = \frac{\langle I_p(t)I_p(t+\tau)\rangle_{\vec{r}}}{\langle I_p(t)\rangle_{\vec{r}}\langle I_p(t+\tau)\rangle_{\vec{r}}} - 1 \tag{1}$$

In Eq. 1, $t$ is the experimental time and $\tau$ is the delay or lag time for the correlation calculation. Images are divided into small region of interests (ROIs), so that spatially resolved dynamics can be obtained. $I_p$ is the intensity at $p$-th pixel, and $\langle ... \rangle_{\vec{r}}$ is the average over the pixels corresponding to a ROI with center position $\vec{r}$, as indicated in Fig. 1(b). For $C$ to be efficiently calculated, the speckle size should roughly match the pixel size [32][a].

## 2. Practical form of the autocorrelation function

At a fixed time $t_0$ and position $\vec{r}_0$, $C$ is a function of the time delay $\tau$ only, $C(\tau)_{\vec{r}_0,t_0}$, whose decay time quantifies the $t$-dependent local dynamics. A convenient functional form that describes well $C(\tau)_{\vec{r}_0,t_0}$ under many experimental conditions is [23, 33]:

$$C(\tau)_{\vec{r}_0,t_0} = A exp\left(-2\epsilon \sqrt{\left(\frac{\tau}{\tau_0}\right)^p + a}\right) + B, \qquad (2)$$

where $\epsilon$ depends on the polarization of the incident and detected light. $a = 3l^*/l_a$ depends on the ratio of the photon transport mean free path $l^*$ and the absorption length $l_a$ (usually, $a \ll 1$). $\tau_0$ is the characteristic decay time, corresponding to the time scale over which $C$ decreases to around $exp(-2\epsilon)$ after normalization (as indicated in Fig. 1(b)), and the scatterers are displaced over a distance of the order of $1/k$. $k$ is the wave vector of the laser light, $k = 2n\pi/\lambda$, with $n$ the refractive index of the material and $\lambda$ the in-vacuum laser wavelength). $B$ is the baseline of the correlation function: ideally, $B=0$, but measurement noise and other artifacts often result in a non-zero baseline [34]. Details of these parameters are described in Supporting Information. Eq. 2 generalizes the usual form of the correlation function in the backscattering geometry [23, 24], with $\left(\frac{\tau}{\tau_0}\right)^p = k^2 \langle \Delta r^2(\tau) \rangle$, with $\langle \Delta r^2(\tau) \rangle$ the mean squared displacement of the scatterers.

Note that by construction Eq. 2 cannot capture complex dynamics, e.g. stemming from the coexistence of two distinct dynamical processes occurring over different time scales. Nonetheless, Eq. 2 is sufficiently flexible to reproduce a variety of microscopic motions of interest, from Brownian motion (corresponding to $p = 1$) to the ballistic motion resulting from an applied deformation at constant rate ($p = 2$). Given the functional form of $C$, $\tau_0$ is also the time scale where the autocorrelation function is most sensitive to a change of the probed

---

[a] In practice, the speckle size may be adjusted by varying the numerical aperture ($f$-number) of the objective mounted on the camera: the higher the $f$-number, the larger the speckle size

time scale. Accordingly, $\tau_0$ is commonly used for quantifying and comparing different dynamical processes[4, 32, 35]. Examples of autocorrelation functions are shown in Fig. 1(b), which illustrates that slower dynamics corresponds to a slower decay rate of $C$ and hence a larger $\tau_0$ value.

## C. Dynamical processes probed by MSDWS

Coexisting dynamic processes yielding multiple modes of decorrelation are commonly reported in DWS measurements [33, 36, 37]. Understanding these dynamics is the key challenge posed by the method. The measured dynamics is often attributed to a single relaxation mechanism, since it is very difficult to reproduce and analyze the form of the autocorrelation function in the general case. However, this requires a careful examination and choice of the system and experimental parameters. Here we briefly review the most common contributions to the dynamics of soft solids and discuss the coupling between them. In particular, we show that in a wide range of practical mechanical testing conditions, MSDWS results can be safely analyzed by considering only the contribution due to the affine deformation field, which greatly simplifies the processing.

### 1. Different dynamics sources in MSDWS measurement

One source of dynamics for scatterers in a soft solid is thermally activated motion. Let us first consider the case of a soft solid whose turbidity is high enough for MSDWS to be viable with no addition of probe particles. Thermal energy results in (overdamped) fluctuations of the structure around its equilibrium position (think, e.g., of the fluctuations of strands in a colloidal gel or of network vibrations in elastomers). The amplitude of these fluctuations depends on the elastic modulus of the material, and is typically on the order of several nanometers for stiff materials up to tens of microns for very weak colloidal gels [38, 39]. Since multiple scattering is sensitive to motion down to the nanometer scale [23, 24], these fluctuations are potentially captured by the correlation function. However, their typical time scale is quite fast, typically 0.001 s or less [37, 39, 40]. On longer time scales, fluctuations are limited by the sample structure [41, 42], so that this fast decorrelation mode normally does not lead to a full decorrelation. Because MSDWS uses a relatively slow detector (340 Hz for our camera), thermal fluctuations do not have a measurable impact on $C$, because they occur out of the experimentally accessible time window. More specifically, they result in overall lower levels of the correlation function measured by MSDWS, independently of $\tau$, which is corrected by the normalization factor *A* in Eq. 2. The situation is similar for samples to which particle probes have been added to enhance scattering, as long as the particles are large enough to be effectively trapped in the sample structure (e.g. if they are larger than the mesh size for gels and elastomers), as is the case for microrheology [38].

An additional source of dynamics in soft solids is the slow evolution of the sample, also referred to as aging [32, 43, 44]. Aging is due to the fact that many soft solids are out-of-equilibrium systems, slowly evolving towards (but often never reaching in accessible time scales) the most thermodynamically stable configuration. In the last two decades, ultraslow relaxations associated with aging have been uncovered in a wide range of systems, from gels to dense, repulsive systems [45-47]. These relaxation modes often lead to ballistic dynamics, which have been ascribed to the progressive release of internal stress stored when forming the sample [48, 49]. Even when measured with techniques such as MDWS that are extremely sensitive to motion on small length scales, the decay time of these modes may be as large as hundreds or thousands of seconds [4] and it further increases with sample age. As we shall see, the decay time associated with an imposed mechanical deformation is typically smaller, on the order of a few seconds at most. Thus, aging dynamics usually do not interfere with the dynamics measured while mechanically driving a soft solid. However, one should always check this point, by running a preliminary measurement on the sample at rest. Should the spontaneous dynamics be fast enough to potentially interfere with those measured under a mechanical drive, it is preferable to let the sample age until the spontaneous dynamics become slow enough.

A third source of dynamics, specific to soft solids under mechanical loading, stems from the deformation field induced by loading. Indeed, light scattering is sensitive to the relative motion of the scatterers. When the displacement is not spatially uniform, scatterers are displaced by a position-dependent amount, which results in the decorrelation of the scattered light. The scatterers displacement contains in general an affine contribution (i.e. the displacement field expected for an ideal, homogeneous elastic body under the same loading conditions) and, possibly, non-affine contributions [50, 51]. We first discuss the affine contribution, assuming that non-affine dynamics may be neglected. Under these conditions, the autocorrelation function can be related to the strain field [52] by:

$$C(\tau) = A exp\left(-2\epsilon\sqrt{3k^2 l^{*2} f[\boldsymbol{U}(\tau)] + a}\right) + B \tag{3}$$

$\boldsymbol{U}(\tau)$ is expressed as the product of $\boldsymbol{D}$ times the time interval by:

$$U_{ij}(\tau, t) = D_{ij}(t)\tau \tag{4}$$

where $\boldsymbol{D}$ denotes the rate of deformation tensor [53, 54] and it is assumed that $\tau$ is short enough for the deformation to occur at constant rate. $\boldsymbol{U}(\tau)$ is a rank-two tensor describing the strain field over the time interval $\tau$, and $f[\boldsymbol{U}(\tau)] = [Tr(\boldsymbol{U})^2 + 2Tr(\boldsymbol{U}^2)]/15$ [20, 52, 55].

By replacing Eq. 4 in Eq. 3 and comparing with the general form, Eq. 2, one finds that for a deformation at a constant rate

$$3k^2 l^{*2} \tau^2 f[\boldsymbol{D}] = \left(\frac{\tau}{\tau_0}\right)^p \tag{5}$$

This implies $p$ = 2 (ballistic dynamics) and

$$\tau_0 = \frac{1}{kl^*\sqrt{3f[\boldsymbol{D}]}}. \tag{6}$$

As an example of simple test geometries, we shall consider uniaxial extension and simple shear and discuss typical time scales for $\tau_0$ due to affine deformation.

For uniaxial extension along the $x$ axis in an incompressible solid with true strain rate $\dot{\lambda}_T$ one has [20]:

$$\boldsymbol{D} = \begin{bmatrix} \dot{\lambda}_T & 0 & 0 \\ 0 & -\frac{\dot{\lambda}_T}{2} & 0 \\ 0 & 0 & -\frac{\dot{\lambda}_T}{2} \end{bmatrix} \tag{7}$$

For simple shear along the $x$ direction at shear rate $\dot{\gamma}$ [56-58]:

$$D = \begin{bmatrix} 0 & \frac{\dot{\gamma}}{2} & 0 \\ \frac{\dot{\gamma}}{2} & 0 & 0 \\ 0 & 0 & 0 \end{bmatrix} \quad (8)$$

Eq. 8 results from decomposing the shear rate tensor in the sum of two terms, corresponding to pure shear and pure rotation, respectively [58]. Recalling that MSDWS is only sensitive to the relative motion of the scatterers, one recognizes that the only term that contributes to the decay of the correlation function is the pure shear term, which corresponds to Eq. 8. Then, for uniaxial extension:

$$\tau_{0,ue} = \sqrt{\frac{5}{3}} \frac{1}{kl^* \dot{\lambda}_T} \quad (9)$$

and for simple shear:

$$\tau_{0,ss} = \sqrt{5} \frac{1}{kl^* \dot{\gamma}} \quad (10)$$

The above discussion shows that the MSDWS relaxation time in the backscattering geometry for a sample undergoing a macroscopic deformation depends on $l^*$, unlike the case of the spontaneous dynamics of a system at rest [23]. Indeed, in multiple scattering experiments $l^*$ sets the scale over which the deformation is measured [35].

As mentioned above, the minimum time delay accessible to MSDWS is about 1 ms (2.9 ms for camera used in this work). This sets a limit on the highest measurable deformation rate, since the relaxation time of the correlation function must be at least a few times the smallest $\tau$, in order to be correctly measured. Using typical parameter values $l^* = 100$ μm, $n=1.5$, $\lambda = 532$ nm, $\tau_{0,min} = 5$ ms, one obtains the following estimates:

$$\dot{\lambda}_{T,max} \approx \frac{7.3 \times 10^{-4}}{\tau_{0,ue\,min}} \approx 0.14 \text{ s}^{-1} \quad (11a)$$

$$\dot{\gamma}_{max} \approx \frac{1.26 \times 10^{-3}}{\tau_{0,ss\,min}} \approx 0.25 \text{ s}^{-1} \quad (11b)$$

To access higher deformation rates, one may reduce $l^*$ by increasing the concentration of tracer particles, or by choosing tracer particles with a higher refractive index contrast with the sample, or by optimizing the size of the tracer particles, since $l^*$ has a non-monotonic behavior with particle size [23]. Note that using a faster camera to decrease the minimum measurable relaxation time $\tau_{0,min}$ may not be a useful strategy, because the relaxation dynamics due to thermal fluctuations discussed above would likely contribute to the decay of the correlation function, making data analysis difficult. The lower limit on the measurable deformation rate is of the order of a few $10^{-6}$ s$^{-1}$, set by the longest relaxation time achievable by MSDWS, which is of the order of a few hundreds of seconds at least. Therefore, the accessible range of deformation rates covers most of the conventional measurement of mechanical behavior in soft solids. Within the accessible strain rate interval, spatially and temporally resolved local strain rates can be directly quantified by MSDWS, provided that $l^*$ is known, e.g. from an independent transmission measurement [23]. $l^*$ may be deduced from Eqs 9 or 10, by imposing a known deformation rate to the sample. Note that the components of the deformation tensor can be directly calculated by MSDWS for simple, ideal deformation geometries. In the more general case (e.g. near to a fracture tip or in more complex geometries), $\tau_0$ only provides the time scale over which $\sqrt{Tr(\mathbf{D^2})}$ reaches $1/kl^*$ (see Eq. 5), with no information on the individual components. In the case where dynamics are only dominated by the affine deformation field, the spatially-resolved strain rate can be quantified by $\tau_0$ obtained by fitting the autocorrelation functions measured in different ROIs. Alternatively, one can also plot the correlation value (C) at a fix time delay $\tau$ to qualitatively visualize and identify spatial heterogeneity, thus obtaining "dynamic activity maps" (DAMs). This approach will be discussed later when analyzing the experiments presented in section V.

A fourth source of dynamics for soft solids loaded beyond their linear regime stems from rearrangements associated with plastic events, as exemplified in Refs. [25-27], which apply MSDWS to the detection of plastic events in granular or polymeric systems. In general, we expect a plastic event occurring at a position $\vec{r}_0$ to have a two-fold impact on the dynamics measured by MSDWS. On the one hand, the local configuration of the sample around $\vec{r}_0$ is modified, e.g. as in T1 or T2 events in foams [59] or in shear transformation zones in dense suspensions [60] or in a bond breaking event in network-forming systems [25]. The size of the region directly modified by the plastic event depends on the system, but it is typically smaller than or comparable to $l^*$. On the other hand, a local plastic event modifies the internal stress state, which in turn sets a strain field across the sample, including far from $\vec{r}_0$. MSDWS is sensitive to displacements down to the nanometer scale, but the spatial mapping of the dynamics is coarse-grained on a length scale $\sim l^*$ or larger (hundreds of μm). Consequently, it is likely that plastic events are not detected *per se*, but rather thanks to the strain field they set throughout the sample.

A final issue to be considered in space-resolved MSDWS measurements coupled to mechanical tests is the rigid displacement of the sample, and hence of the speckle pattern. The rigid displacement causes a loss of the correlation in the autocorrelation function[20, 22, 61]. As a limiting case, consider a sample undergoing a pure translation. Due to the imaging geometry, the (otherwise frozen) speckle pattern is displaced with respect to the detector as the sample translates, such that the intensity of any given pixel fluctuates as bright and dark speckles scroll in front of it. For a sample undergoing tensile or shear strain, a similar effect occurs and the measured loss of correlation is in general due to both a rigid displacement of the speckle pattern and its evolution associated with the change of the relative position of the scatterers, stemming from the sample deformation. Schemes have been proposed to disentangle the two contributions [61]. Since these methods are computational demanding, it is interesting to discuss under what conditions the contribution of the rigid displacement may be safely neglected.

## 2. Coupling between rigid displacement and deformation during mechanical testing

The contribution of the rigid displacement is negligible when $\tau_{speckle}$, the time it takes for a ROI to travel a speckle size due to rigid displacement, is much longer than $\tau_{0,\text{def}}$, the relaxation time due to the relative motion of scatterers associated to the macroscopic deformation, see e.g. Eqs 9 and 10 for the case of extensional and shear tests. For the former, one has

$$\tau_{speckle} = \frac{l_{speckle}}{v} = \text{FOV}/(Nv), \tag{12}$$

where $l_{speckle}$ is the speckle size, which can be measured by calculating the spatial autocorrelation function of one single speckle image [62], FOV is the size of the field of view and $v$ is the speckle displacement velocity. The last equality stems from the speckle size being a fraction $1/N$ of the FOV. This formulation is used in the following, since the field of view is a key parameter of interest in experiments, while $N$ can be easily controlled by closing (resp., opening) the diaphragm of the objective lens, which results in smaller (resp., larger) $N$ values.

The cameras used in MSDWS produce images with a linear size of several hundreds to a few thousands of pixels, while the speckle size ranges from a fraction of a pixel to several pixels: thus, $N$ ranges from a few hundreds to a few thousands.

We first consider an extensional test with only one clamp moving, and the other one fixed (commonly to a force sensor). A popular geometry consists in imaging the whole sample, which allows to obtain full information on the sample response and to check for slippage at both clamps. Under these conditions, FOV = $L$, $L$ being the sample length. The largest velocity occurs at the moving clamp: $v = \dot{\lambda}_N L$, where $\dot{\lambda}_N$ is the nominal strain rate. Here we simply consider $\dot{\lambda}_N = \dot{\lambda}_T$ to estimate the approximately scale for comparison. Inserting this expression in Eq. 12 and using Eq. 9, we find the following lower bound

$$\frac{\tau_{speckle}}{\tau_{0,ue}} \geq \sqrt{\frac{3}{5}\frac{kl^*}{N}} \qquad (13)$$

Using the same parameters as above ($l^* = 100$ µm, $n = 1.5$, $\lambda = 532$ nm), one finds $\tau_{speckle}/\tau_{0,ue} \approx 1370/N$. Remarkably, the ratio of the two characteristic times does not depend on the strain rate and may be tuned to be either smaller or larger than unity by varying the speckle size, which in turn controls $N$. For example, we will show in section IV that in the experiments on a stretched PDMS elastomer presented here the speckle size was such that $kl^* = 7260$, $N = 683$ and $\tau_{speckle} = 8.2\tau_{0,ue}$. Under these conditions, the speckle pattern fluctuates (due to the affine displacement field) much faster than it translates (due to the displacement associated to elongation), such that no correction needs to be implemented. On the contrary in Ref. [20] the speckle size was designed to be smaller than the pixel size, such that $\tau_{speckle} < \tau_{0,ue}$ and the strain rate field was inferred from the displacement field, measured using the computationally demanding method of Ref. [61].

For a shear test, we consider a sample of thickness $d$ undergoing simple shear. We recall that MSDWS in the backscattering geometry is not sensitive to displacements in the full depth of the sample, but rather in a layer of thickness $\sim 10 l^*$ on the side of the illumination and collection optics[63]. Assuming that the sample is sheared at a rate $\dot{\gamma}$ while keeping fixed one of its surfaces, one has to distinguish the two cases where the sample is illuminated on the fixed side or on the mobile side. For a sample illuminated on the fixed side and assuming $d > 10 l^*$, the displacement velocity averaged over the thickness of the probed layer is $v = 5 l^* \dot{\gamma}$. Inserting this expression in Eq. 12 and using Eq. 10, one finds the lower bound

$$\frac{\tau_{speckle}}{\tau_{0,ss}} \geq \frac{FOV\, k}{10\, N}. \tag{14}$$

Note that the right-hand side of Eq. 14 is particularly simple, since it does not depend on strain rate nor on $l^*$. With the typical values used above, $k \sim 1.7 \times 10^7$ m$^{-1}$, such that for usual geometries (FOV $\geq$ 1 cm and $N \sim 1000$) $\tau_{speckle}/\tau_{0,ue} \geq 20$. Thus, for simple shear while imaging on the fixed side of the sample the speckles fluctuate much faster that they are displaced and no correction is required. This will be the situation for the experiments on the adhesive debonding reported in section V.

If a sample of thickness $d$ undergoing shear is imaged from the moving side, the velocity averaged over the layer probed by MSDWS is $v = \dot{\gamma}(d - 5l^*)$ (assuming again that $d > 10 l^*$) In this case, the lower bound for the ratio of the displacement and fluctuation times becomes

$$\frac{\tau_{speckle}}{\tau_{0,ss}} \geq \frac{FOV\, k l^*}{\sqrt{5}\, (d-5l^*) N}, \tag{15}$$

which does not depend on the strain rate, but does depend on $l^*$. Inserting in Eq. 15 typical values for the various parameters (FOV = 1cm, $l^*$ = 100 μm, $d$ = 1 mm), one finds $\tau_{speckle}/\tau_{0,ue} \geq 16$, showing that the correction of rigid displacement may be safely neglected.

To summarize this section, we have shown that in the typical range of time delays probed by MSDWS coupled to mechanical tests (1 ms < $\tau$ < 100 s), the relaxation time of the intensity correlation function is directly related to the (local) deformation rate. The latter may be extracted from the MSDWS data provided that $l^*$ is known and that the deformation geometry may be approximated by a simple geometry such as extension or shear. While in principle correcting the speckle images for rigid displacement may be necessary, we find that for typical experimental parameters this correction is not necessary, apart for tensile tests when the speckle size is small enough for thousands of speckles to fit in the (linear) size of the image.

## D. Comparison of MSDWS and DIC

As MSDWS and DIC can provide similar information on the deformation rate spatial distribution, it is worth to quantitatively compare the two methods. Considering that the correlation function is sufficiently informative when $\tau \approx 0.02\tau_0$ and $C(\tau) \approx 0.93$, this gives a strain accuracy for MSDWS ($\approx f[U(0.02\tau_0)]$) of the order of $10^{-6}$. This is much more accurate than DIC, whose strain resolution is limited to the scale of $10^{-3}$ in practical experimental conditions[64, 65]. Another advantage of MSDWS is the fact that the method only requires the material to be turbid (which may also be achieved by adding a small amount of highly scattering particles), with no need of surface treatment. More importantly, MSDWS directly provides the magnitude of the deformation rate, which can be more useful for the examination of subtle dynamic changes, compared to conventional strain measurements.

## III. Material and methods

### A. Materials

#### 1. Elastomer samples for tensile tests

For the tensile tests, a Sylgard 184 Polydimethylsiloxane (PDMS) elastomer was prepared with ratio of PDMS base and curing agent of 10:1. Titanium dioxide ($TiO_2$) nanoparticles (diameter 250 nm measured by Dynamic light scattering (DLS)) with wt% of 0.25 % were added before curing, to enhance the sample turbidity. $TiO_2$ nanoparticles were first mixed with the PDMS base and manually stirred. The suspension was then sonicated for 10 min to thoroughly disperse the particles and put in vacuum to remove bubbles formed during stirring and sonication. The suspension and curing agent were mixed and poured into a mold, then cured at 90 °C for 12 h. A sheet of elastomer of 4 mm thickness was thus obtained. The transport mean free path $l^*$ of the PDMS elastomer was measured in a suspension containing $TiO_2$ nanoparticles with the same volume concentration as in the elastomers. The autocorrelation function from the suspension can be calculated from the known viscosity of water and matched to the experimental one, using $l^*$ as the only fitting parameter [23]. For a PDMS wt% = 0.25 %, we find $l^* \approx 430$ µm. Note that here the dependence of scattering cross section on the difference between the refractive index of water and that of PDMS is ignored, which may lead to a small deviation of the calculated $l^*$, which will be discussed later.

#### 2. pressure sensitive adhesive tapes for debonding tests

Commercially available Polyacrylate pressure sensitive adhesive tapes were used to investigate debonding. The tapes are made of a copolymer of 2-ethylhexyl acrylate (88 % by weight) and acrylic acid (12 % by weight), filled with hollow glass microspheres (12 % by weight, Type: Q-Cel® 5020 from Potters, with mean size of 60 µm). The copolymer was synthesized by UV-polymerization using Irgacure 184 (0.4 % by weight) as photoinitiator and 1,6-Hexanediol diacrylate (0,15 % by weight) as crosslinker. The thickness is 1 mm and the tapes are cut to different sizes for measurement. To determine the photon mean transport path $l^*$, we used the method discussed at the end of section IV (measurement of the decay time of $C(\tau)$ while applying a tensile strain at a known deformation rate), finding $kl^*$ =

$3.3 \times 10^3$, corresponding to $l^* \approx 190$ μm, assuming $n= 1.47$ (the refractive index of 2-ethylhexyl acrylate).

## B. Experimental set-up

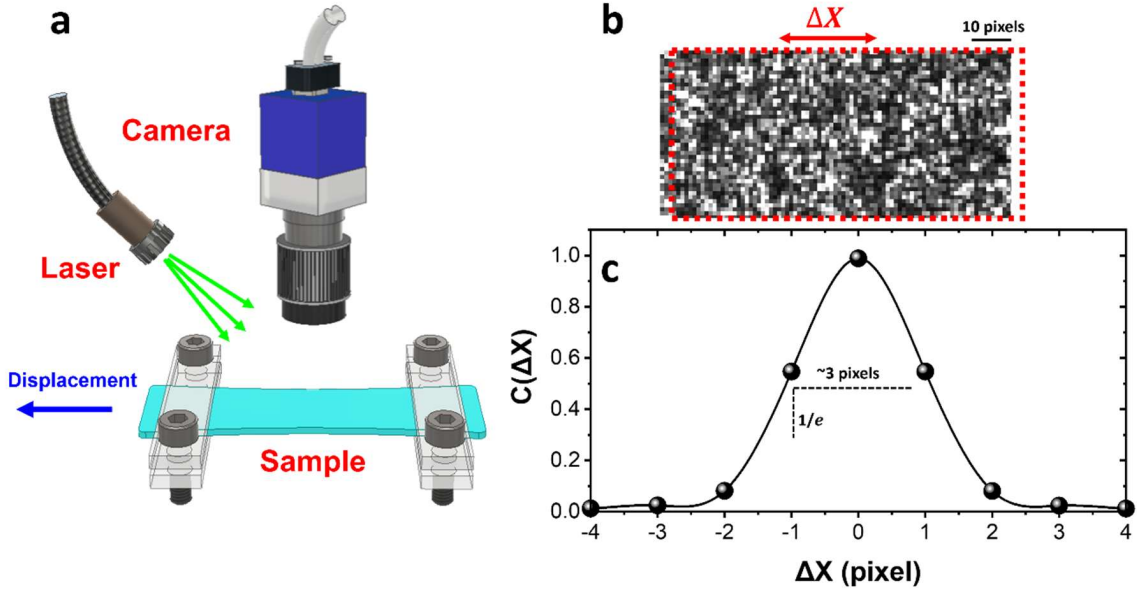

*Fig. 2 (a) Experimental set-up for the uniaxial extension experiment, combined with the MSDWS measurement. (b) A typical portion of a speckle image and its displaced position. (c) Spatial autocorrelation function of image in (b), from which the size of the speckle is estimated at $C = 1/e$ to be about 3 pixels.*

### 1. Uniaxial extention test set-up

The uniaxial extension test of the elastomer was performed with a custom set-up, see Fig. 2(a) (a detailed image of the set-up is shown in Fig. S1 of the Supporting Information.). The sample was fixed by two clamps, one of which is connected to a Newport translational stage with a step precision of 500 nm. The nominal stretch $\lambda_N$ is defined as the ratio of the deformed length over the original length. The whole sample surface was illuminated by an expanded laser beam (Excelsior-532-300-CDRH, wavelength 532 nm, power 300 mW) with incident angle around 45° and diameter around 6 cm at the surface of the sample. The sample/laser distance is about 20 cm and the camera/sample distance (via a mirror) is around 25 cm. Images of the sample surface formed by the backscattered light were collected simultaneously to the nominal displacement imposed by the motor, using a CMOS camera (BASLER acA2000-340km, sensor size 11.3 mm × 6 mm, pixel bit depth 8 bits). The camera was equipped with an objective lens (Thorlabs, MVL50TM23) with focal length of 50 mm.

Fig. 2(b) shows a typical portion of the CMOS image where the presence of speckles can be detected. Spatial autocorrelation (using spatial displacement $\Delta X$ instead of time interval $\tau$ in Eq. 1 for the correlation, as seen in Fig. 2(b)) is calculated and shown in Fig. 2(c). The size of speckle is estimated by the full width at $C = 1/e$, around 3 pixels. The image size in the horizontal direction (also the deformation direction) is 2048 pixels, hence $N$ = 2048/3 = 683. Inserting this value and $n$ = 1.43 for PDMS[66] and $l^*$ = 430 μm into Eq. 13, one has $\tau_{speckle} = 8.2\tau_{0,ue}$, so that there is no need for the correction of rigid displacement. We used digital image correlation (DIC) to check that no slippage occurred at the clamps, testing a stretched PDMS sample with fixed $\lambda_N = 1.15$ for more than 200 s.

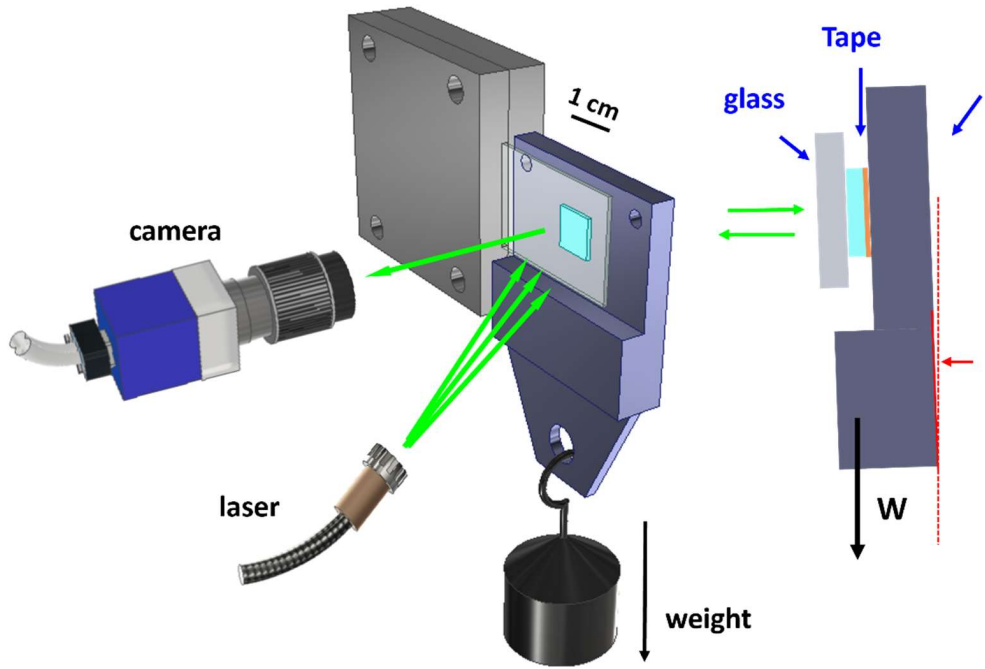

*Fig. 3 Experimental set-up for the measurement of adhesive debonding under single-lap shear.*

## 2. Adhesive tape debonding test set-up

The debonding process of a pressure sensitive adhesive tape in single-lap shear[67, 68] (tape fixed between two rigid substrates and exposed to a translational shear stress) was measured with the set-up shown in Fig. 3. The camera and laser configuration are similar to that in the set-up for uniaxial extension. The sample/laser distance is around 20 cm and the camera/sample distance is around 10 cm. The angle of incidence of the laser beam on the sample is about 45°.

First, the tape was glued to an aluminum plate using instant glue (Loctite 406) to ensure adhesion stronger than at the debonding interface. Then, a glass microscope slide with a thickness of 3 mm was pressed to the other side of the tape, using a weight of 0.3 N for 10 min. A shear stress was imposed to the tape by holding the glass slide fixed in a vertical plane, while hanging a weight to the aluminum plate. Rectangular samples were used, with different sizes as shown in Table 1. The position of the loading axis was set to be around the midplane of the tape, to avoid torque on the tape, which may result in strong peel stresses and cleavage in the bulk region. A small angle around 2.5° is introduced between the aluminum plate and the vertical axis, so that a small component of the load provides compression on the tape, to avoid possible normal tension that strongly changes the debonding conditions. In this geometry, the tape is under simple shear conditions and adhesive debonding is studied under creep deformation. The weight $W$ was adjustable from 1.5 N (the weight of the aluminum plate alone) to 21.5 N. Macroscopic shear stress during creep is calculated by $\sigma_{macro} = W/A$, where $A$ is the area of the glass-tape interface, typically of the order of a fraction of cm² (Table 1). The laser illumination and the imaged surface is on the side of the glass-tape interface, which is static until debonding.

To estimate whether any correction may be needed, we use a slightly modified version of Eq. 14, because here $d/l^* \sim 5$ and MSDWS probes the dynamics across the full thickness of the sample. The average velocity of the measured layer is $v = 2.5 l^* \dot{\gamma}$ and $\tau_{speckle}/\tau_{0,ue} = FOV\, k/5N = 107$, where we have used $k = 1.67 \times 10^7$ m$^{-1}$, $N = 250$, FOV $\approx 0.8$ cm for the biggest sample (8.4 mm × 7.6 mm, where $N$ is maximal). Since $\tau_{speckle} \gg \tau_{0,ss}$, no correction was necessary for the debonding tests concerning rigid displacement.

| $\sigma_{macro}$ (MPa) | Tape size (length × width, mm) | $W$ (N) |
| --- | --- | --- |

| 0.26 | 8.4 × 7.6 | 16.5 |
| 0.49 | 5.2 × 5.5 | 13.5 |
| 0.62 | 5.2 × 5.1 | 16.5 |
| 0.72 | 4.5 × 3.5 | 11.5 |
| 0.88 | 5.1 × 3.6 | 16.5 |

*Table 1 Parameters for debonding tests under shearing.*

The samples were illuminated by a diode green laser with $\lambda$ = 532 nm (wavelength 532 nm, power 300 mW). Speckle images in the backscattering geometry were collected using a CMOS camera (BASLER acA2000-340km). To optimize the time resolution and speckle quality, the exposure time was set to 0.03 s. The strong spatial fluctuations of intensity due to the speckled appearance of individual images makes it difficult to obtain morphological information on the surface, such as evidence of cavitation [69] and fibrillation [70] that typically occur during debonding. However, the morphological information can be extracted by averaging several speckle images acquired over a time interval longer than $\tau_0$, such that the intensity fluctuations average out. The displacement $L$ of the aluminum plate can be measured as a function of time by tracking a reference point on the aluminum substrate and the macroscopic shear strain is defined as $\gamma_{macro} = L/d$, where $d$ is the sample thickness, 1 mm for all measurements.

## C. Software and data processing

Images are collected by commercial software Norpix Streampix 8 and data are processed by self-developed program in Python.

# IV. Method validation: uniaxial extension of PDMS

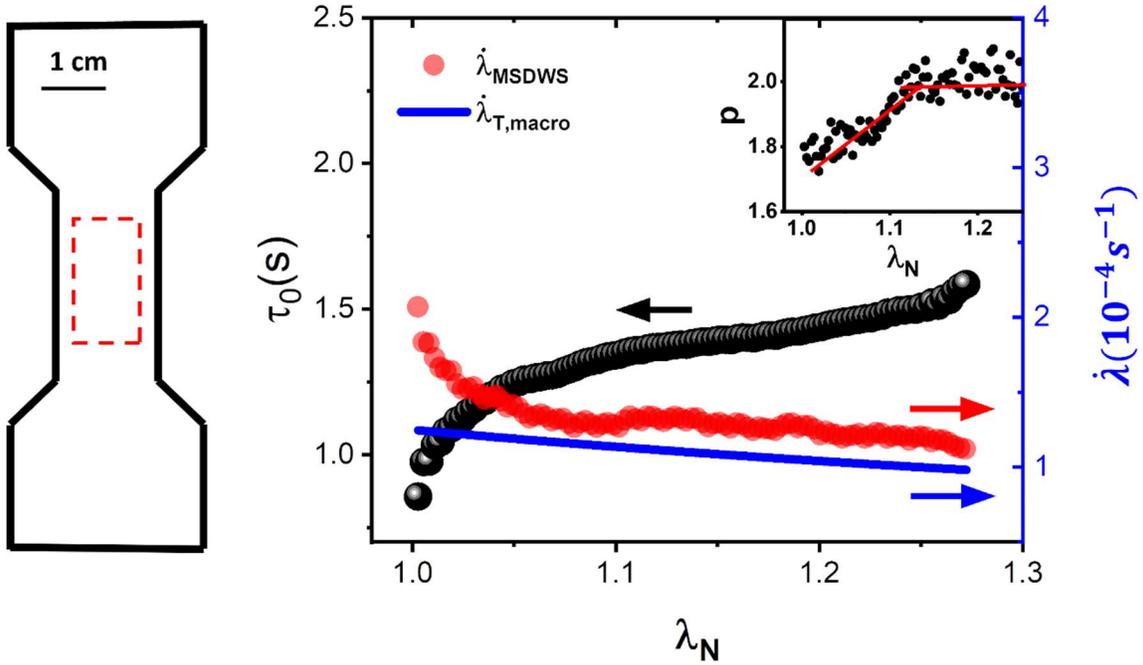

*Fig. 4 Measurement of the deformation rate in the central region of a dog-bone PDMS sample, schematically shown on the left. Graph: relaxation time $\tau_0$ measured by MSDWS (black circles, left axis) and corresponding experimental (red) and nominal (blue) strain rate $\dot{\lambda}$ (right axis), as a function of $\lambda_N$. Inset: exponent p as a function of $\lambda_N$.*

## A. Averaged strain rate measurement

We first validate MSDWS applied to mechanical tests in uniaxial extension of a dog-bone PDMS sample to which a nominal strain rate $\dot{\lambda}_N = 1.25 \times 10^{-4}$ s$^{-1}$ is applied. The autocorrelation function is calculated for a ROI located in the center of the sample (with size of 1 cm × 2 cm, indicated by the dashed rectangle in Fig. 4). The autocorrelation function is fitted to Eq. 2, using $a$, $p$ and $\tau_0$ as fitting parameters. We find that $a$ is around 0.015 from fittings at different $\lambda_N$, so that the absorption is very low. Note that $a$ should be essentially constant throughout the experiment but its fitted value strongly depends on data quality. In the following fittings, we used a constant value $a$ = 0.015. An example of fitting is shown in Fig. S2 in supporting information. The relaxation time $\tau_0$ is plotted as a function of the nominal strain $\lambda_N$ in Fig. 4. We find that $\tau_0$ is slightly smaller than 1 s at the beginning of the test and increases up to about 1.5 s with growing strain.

Assuming that the observed dynamics are only due to the imposed strain, the strain rate $\dot{\lambda}_{MSDWS}$ is calculated from $\tau_0$ using Eq. 9 with $l^* = 430$ μm and plotted in Fig.4 (red circles, right axis). A sharp decrease is detected in the initial stage of the test ($\lambda_N < 1.1$), after which $\dot{\lambda}_{T,MSDWS}$ keeps decreasing, but more slowly. Due to the shrinkage of the cross-sectional area upon stretching, the macroscopic true strain rate may be calculated from the macroscopic nominal strain and macroscopic nominal strain rate:

$$\dot{\lambda}_{T,macro} = \frac{\dot{\lambda}_N}{\lambda_N} \qquad (16)$$

The value of $\dot{\lambda}_{T,macro}$ is plotted in Fig.4 as blue line (right axis). We found that the macroscopic values obtained from Eq. 16 match well with $\dot{\lambda}_{MSDWS}$ for $\lambda_N > 1.1$, with around 15 % of deviation. Concomitantly, the exponent $p$ (inset of Fig. 4) increases from 1.8 and plateaus to $p = 2$ for $\lambda_N > 1.1$. This suggests that there is an initial regime where other relaxation mechanisms, in addition to affine displacements, contribute to the decorrelation of the autocorrelation function. We speculate that these additional relaxation mechanisms may be due to heterogeneity of the local environment of the probe particles and from residual stresses generated when clamping the sample. Indeed, in preliminary tests on the same material with no extension, we observed enhanced dynamics upon clamping, which slowly relaxed over several hours [71]. The contribution of these additional mechanisms becomes negligible beyond $\lambda_N = 1.1$, where the evolution of $\dot{\lambda}_T$ as measured macroscopically and by MSDWS match very well. The 15% systematic offset between the microscopic and macroscopic strain rates is likely to stem from a difference between the values of $l^*$ in the PDMS sample and as obtained from measurements of the dynamics of aqueous suspensions of the same nanoparticles, respectively. Indeed, the proportionality of the two curves seen in Fig. 4(b) strongly suggests that the value of $l^*$ used to analyze the MSDWS data via Eq. (9) is slightly underestimated. This is consistent with the fact that the mismatch between the refractive index of the $TiO_2$ particles and that of the medium in which they are dispersed is larger in water than in PDMS, which would lead to a smaller $l^*$ in water as compared to that in PDMS[23]. Note that the proportionality established in Fig. 4b implies that the actual value of *l\** may be obtained by matching the deformation rate measured by MSDW the macroscopic one.

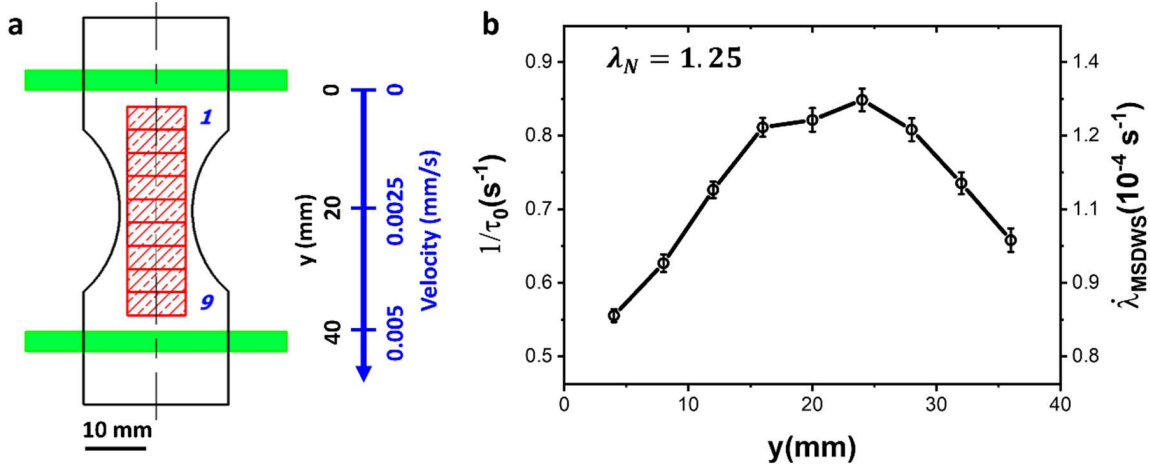

*Fig. 5 (a) PDMS dog bone sample with a varying width along the stretching direction. The green rectangles indicate the clamp positions. A tensile strain is imposed by moving the bottom clamp at a velocity of 0.005 mm/s. The red hatched rectangles show the location of the 9 ROIs used for the data analysis. (b) $v_0$ and the corresponding local strain rate measured by MSDWS as a function of the ROIs position.*

## B. Spatially-revolved strain rate measurement

We argued in section IIIB that no correction for the rigid displacement of the sample was required in the experiments on PDMS reported here. To validate experimentally this result, the contribution of the rigid displacement is investigated using a customized dog bone sample, whose width along the extension direction varies, as shown in Fig. 5(a). A tensile strain is imposed by moving one clamp at a velocity of 0.005 mm/s, such that the displacement velocity is proportional to the distance $z$ from the fixed clamp. The sample length is 40 mm and the nominal strain rate is $\dot{\lambda}_N = 1.25 \times 10^{-4}$ s$^{-1}$. Leveraging on the space resolution afforded by MSDWS, we measure $\tau_0$ for nine ROIs, at different positions as shown in Fig. 5(a). The corresponding $\dot{\lambda}_{MSDWS}$ values are calculated using Eq. 9 and are shown in Fig. 5(b). Overall, $\dot{\lambda}_{MSDWS}$ is of the same order of its macroscopic counterpart, $\dot{\lambda}_T = 10^{-4}$ s$^{-1}$, but inhomogeneous along $z$. $\dot{\lambda}_{MSDWS}$ peaks around $z = 22$ mm, close to the middle part of the sample, where the width is smallest, showing enhanced deformation rates as expected due to strain localization. Crucially, $\dot{\lambda}_{MSDWS}(z)$ is nearly symmetric around the middle of the sample, reflecting the symmetry of the sample shape. If the rigid displacement sample would contribute significantly to the decay of $C(\tau)$, this symmetry would be broken and the decay would be faster at larger $z$, where the displacement velocity is higher. Thus, the test on the dog bone sample with variable width confirms that here the dynamics is dominated by the relative motion of the particles, so that only the strain rate is measured, regardless of the displacement velocity.

The experiments on PDMS under uniaxial extension confirm that the spatially-revolved strain rate can be effectively measured by MSDWS. Furthermore, the combination of MSDWS and uniaxial extension provides a simple way to estimate $l^*$ for a (solid) sample with unknown scattering properties, by tuning $l^*$ in order to match the MSDWS strain rate to the macroscopic one. This provides a simple alternative to optical methods to measure $l^*$, which are often quite delicate to implement, especially for solid samples [20, 22]. We shall use this method to determine $l^*$ for the adhesive tapes discussed in the next section.

# V. Calibration of MSDWS measurement of adhesive debonding

## A. Characterizations of kinetics of debonding

MSDWS is applied here to measure the spatially-resoved shear deformation rate at the interface between a soft adhesive film under fixed shear stress $\sigma_{macro}$ and a glass substrate during adhesive debonding, with the set-up shown in Fig. 3. First, for different $\sigma_{macro}$, the nominal (macroscopic) shear strain $\gamma_{macro}(t)$ is measured from the macroscopic displacement in the vertical direction of the aluminum substrate and shown in Fig. 6(a). $\gamma_{macro}(t)$ as a function of time exhibits a typical creep behavior for all $\sigma_{macro}$. A rapid increase in $\gamma_{macro}$ right after loading is observed, followed by a steady growth of $\gamma_{macro}$, exhibiting a power law behavior with a constant exponent around 0.25. The results are in agreement with previous work where creep behavior was studied in the linear viscoelastic regime [72-74]. Finally, a sudden increase of $\gamma_{macro}(t)$ is detected at a critical value of $t$, corresponding to the full debonding. The shear strain where debonding occurs $\gamma_b \sim 3$ (shear strain at break) is almost independent of $\sigma_{macro}$, consistent with a strong strain hardening typical of crosslinked adhesives[75].

The values of time when $\gamma_{macro}$ reaches 1, $t_{\gamma=1}$, are plotted as a function of $\sigma_{macro}$ in Fig. 6(b) to compare the debonding kinetics. The debonding time itself is not used to compare debonding kinetics as it can depend strongly on the edge effects at the interface. We found that $t_{\gamma=1}$ decreases with $\sigma_{macro}$ following power law with exponent around -6, indicating the strong dependence on applied load of the creep failure kinetics. This type of experiment is classically carried out in industry and the only time measured is the failure time. Yet the debonding mechanisms are difficult to investigate and so are the heterogeneities in stress. A recent study on a similar shear failure geometry combining a fracture mechanics approach of shear failure and a finite strain modeling, showed clearly the heterogeneity of the load along the bonded surface[76, 77].

We now focus on the spatial mapping of the deformation process on the interface with the glass. Images of the interface at different times after loading are displayed in Fig. 6(c) for $\sigma_{macro}$ = 0.26 MPa. These images are obtained by averaging 25 speckle images acquired over a time interval larger then $\tau_0$ (the time interval is adjusted during the various experiments to optimize the time resolution and image quality), in order to smooth the spatial fluctuations of intensity due to the speckles and make various morphologies visible. As seen from the enlarged images, bottom row of Fig. 6(c), signs of fibrillation are visible on the top edge of the sample, while they are not visible on the bottom edge. Images are enhanced by averaging over a longer period, with interval time longer than $\tau_0$, so that different images for averaging are uncorrelated. With increasing $\gamma_{macro}$, fibrils develop further and can partially detach from the substrate. After 25 h of loading, 2 h before full debonding and with enhanced contrast, most of the top edge is still attached to the glass substrate with no substantial debonding. At the center of the sample, no particular morphological change is observed from these images.

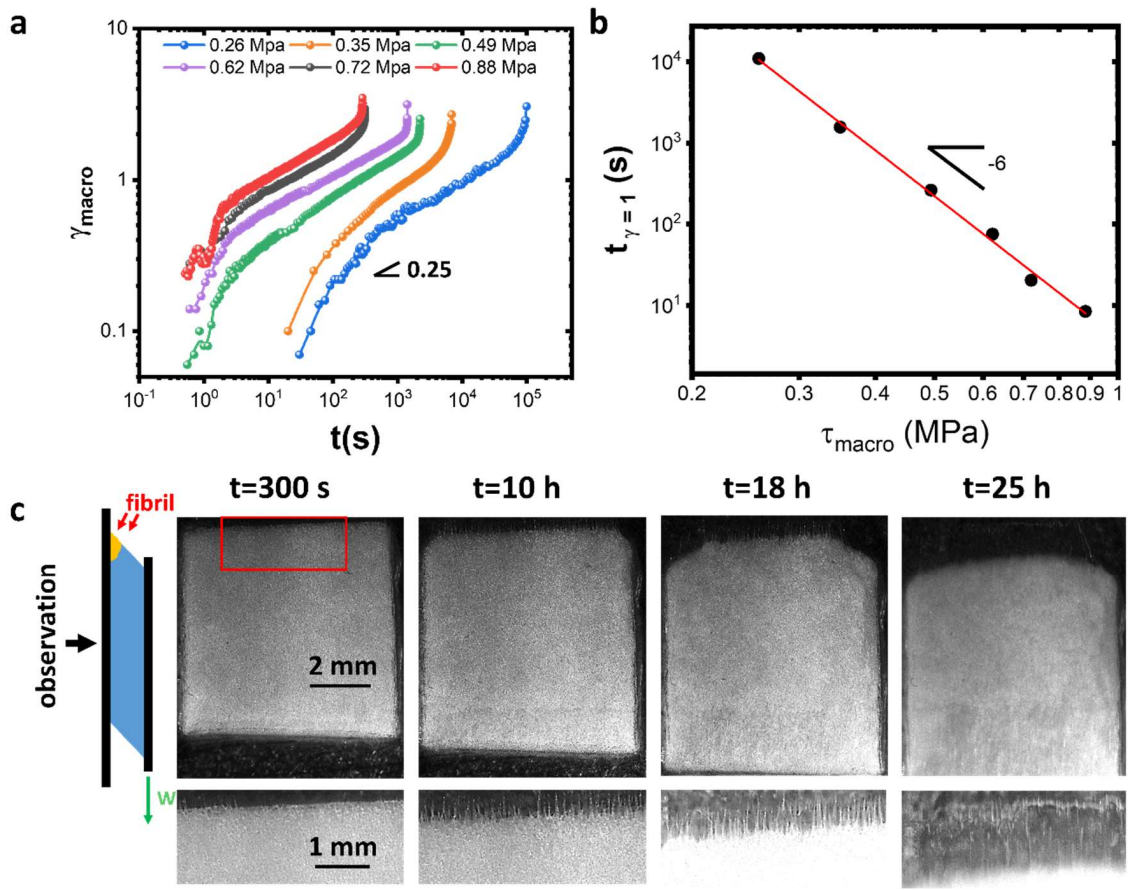

Fig. 6 (a) $\gamma(t)$ as a function of time under different loading shear stresses $\tau_{macro}$. (b) Time where $\gamma_{macro}$ reaches unity, $t_{\gamma=1}$, as a function of $\tau_{macro}$. The data follow a power law with an exponent

*of -6, indicated by the red line. (c) Interface images at different times for $\sigma_{macro}$ = 0.26 MPa. Inset: enlarged images from the region indicated by the red rectangle.*

In previous studies of deformation during adhesive debonding in peel or probe tack tests, it was not possible to measure the in-plane deformation distribution of the debonding interface since the interface is static and bulk deformation is hard to visualize[78-80]. In experiments, debonding was studied by inspecting the sample from the side[81, 82], which does not allow for the full characterization of the large-shear edge region. In principle, DIC can also be applied in this case to measure if there is a slippage at the interface between the substrate and the tape, but the measurement of the deformation field requires speckles that are normally prepared by spray painting, which will largely change the adhesive strength on the interface. By contrast, MSDWS allows for measuring the deformation rate of the adhesive tape over a layer of thickness several $l^*$ without changing its mechanical properties. For our sample, for which $d \approx 5l^*$, this corresponds to the full thickness of the tape, for which we can thus visualize and quantify in-plane heterogeneity in the mechanical behavior of the whole sample.

## B. calibration of deformation rate

In the central region of the tape, where shear localization and peeling are not dominating, we expect the deformation measured by MSDWS to match the macroscopic strain inferred from the aluminum plate displacement, provided that the deformation is homogeneous through the thickness. We compare the MSDWS strain rate and the macroscopic one for a sample with $\sigma_{macro}$ = 0.62 MPa, from the beginning of loading up to $t = 1404\ s$, when debonding occurs. The MSDWS shear strain rate $\dot{\gamma}_{MSDWS}$ is calculated from the fitted relaxation time $\tau_{0,ss}$ of $C(\tau)$ using Eq. 10, with $kl^*$ = 3.3 × $10^3$ independently measured in an uniaxial extension test on the bare tape (see Fig. S3 in the supporting information). The MSDWS data are obtained by averaging $C(\tau)$ in a square region of size 1.5 mm × 1.5 mm located in the center of the sample, as indicated in Fig. 7. The macroscopic strain rate is obtained by numerically differentiating $\gamma_{macro}$ after smoothing the data with a sliding window of 60 s, to reduce noise. Fig. 7 shows that $\dot{\gamma}_{MSDWS}$ matches well the overall trend of of $\dot{\gamma}_{macro}$, although the MSDWS strain rate is systematically larger than the macroscopic one by around 20 %. Such a discrepancy may be due to uncertainties in the sample thickness $d$ used to calculate the macroscopic strain rate or in the value of $kl^*$ obtained from the uniaxial extension test. Furthermore, we analyze the MSDWS data for both the uniaxial extension test and under shear with the formalism of section IIB, which applies to semi-infinite samples, while for our tape samples the condition $d \gg l^*$ is not met. Since the ratio between the microscopic and macroscopic strain rates is close to unity and constant throughout the test, this difference has no impact on the discussion that will follow. We emphasize that the strain rate inferred from MSDWS has a much lower noise than the macroscopic data. MSDWS directly measures strain rate with a superior sensitivity, while macroscopic strain data need to be differentiated, which is a notorious source of experimental noise.

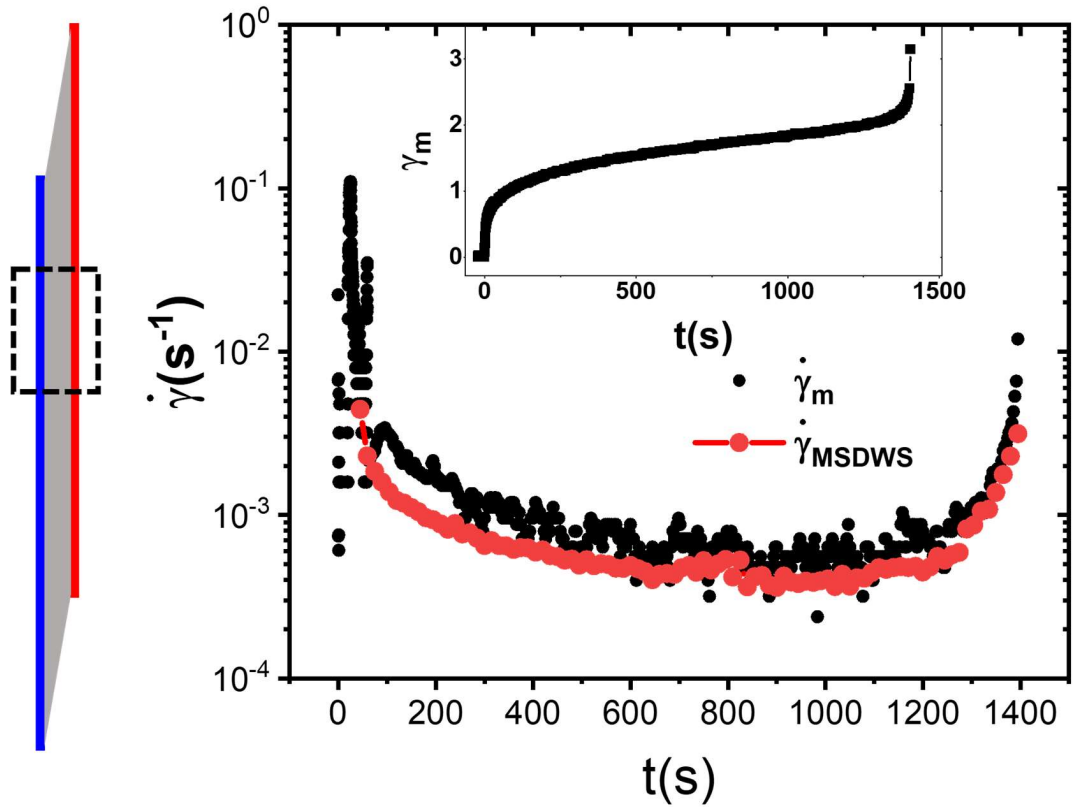

*Fig. 7 Sketch on the left: side view of the sheared adhesive tape, the dashed rectangle indicates the location of the ROI for which the MSDWS data shown on the right are collected. Right: $\dot{\gamma}_{MSDWS}$ and $\dot{\gamma}_{macro}$ as a function of time for $\sigma_{macro}$ = 0.62 MPa. Inset: $\gamma_{macro}$ for $\sigma_{macro}$ = 0.62 MPa.*

# VI. Heterogeneous deformation during adhesive debonding

## A. Strain rate maps

To further characterize the spatial distribution of the strain rate, $\dot{\gamma}_{MSDWS}$ maps of the whole interface at different times are shown in Fig. 8 (a), for the case of $\sigma_{macro}$ = 0.62 MPa. The maps are obtained from spatially resolved autocorrelation functions, by dividing the imaged interface in ROIs with size 0.3 mm × 0.3 mm. The area of the tape that initially adheres to the glass slide is indicated by the red rectangle. White region in the maps corresponds to the debonding region. Higher $\dot{\gamma}_{MSDWS}$ values are detected at the edges, around a central area that bears most of the applied stress. These areas with a higher deformation rate propagate towards the inner region as creep proceeds and debonding sets in. It has been reported that during shear of a single-lap joint, strain and stress at the interface are relatively uniform along the tape, but both shear and peel stress peak at the edge [83]. Previously, this behavior could only be predicted theoretically or investigated numerically. By contrast, MSDWS reveals the spatial distribution of the deformation rate in the adhesion plane. Furthermore, the boundaries of the central area that bears most of the stress can be assessed more reliably than by looking at the emergence of morphological features, as in Fig. 6(c). Note that high values of $\dot{\gamma}$ originate here from both localized shear and peel at the edge, which cannot be distinguished from each other. From about 150 s before global debonding, the local strain rate at the top and bottom edges are significantly enhanced, up to about twice the macroscopic strain rate.

We now turn to a more detailed analysis and quantify the observations in Fig. 8(a). We calculate the profile of the local strain as a function of the distance $y$ from the top edge, by averaging $\dot{\gamma}_{MSDWS}$ over the width of the hatched region (1 mm), shown in Fig. 8(a). We show in Fig. 8(b) profiles of the local shear rate (curves are shifted vertically for better comparison. The value of the local MSDWS strain rate is normalized by the macroscopic scale in each point in Fig. 8(b)). When approaching the debonding time, regions exhibiting high deformation rates further penetrate towards the central region of the sample and $\dot{\gamma}_{MSDWS}/\dot{\gamma}_{macro}$ can locally reach values as high as 2. Crucially, this marked growth of the local strain rate occurs at a time when $\dot{\gamma}_{macro}$ is still stable and the sample is far from full debonding ($\gamma_{macro}$ around

1.7 while $\gamma_b \approx 3$), which occurs at $t$ = 1404 s. These results demonstrate the superior sensitivity and richness of information of spatially-resolved MSDWS, which allow one to detect subtle changes in the deformation rate, hundreds of seconds before the emergence of any macroscopic precursor of debonding.

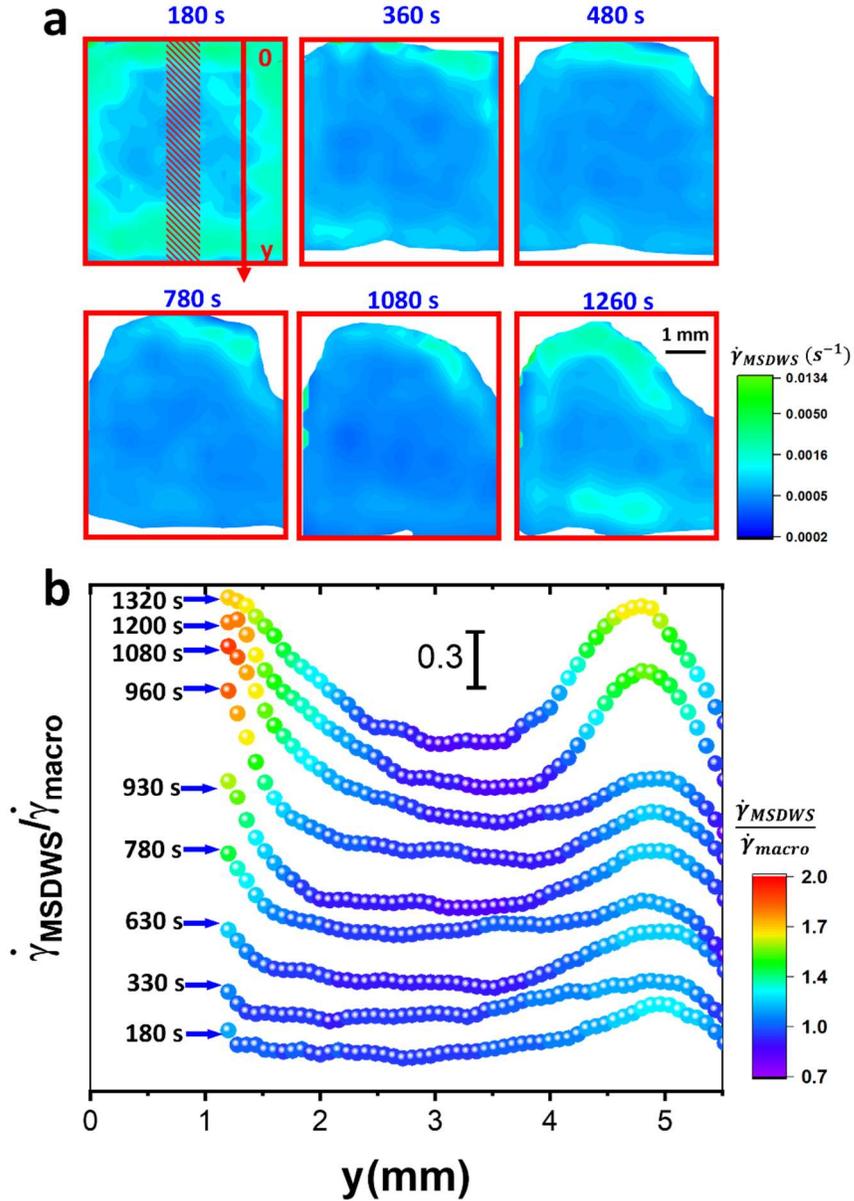

Fig. 8 (a) $\dot{\gamma}_{MSDWS}$ maps at different times $t$ after loading and adhesive tape under shear with $\sigma_{macro}$ = 0.62 MPa (same test as in Fig. 7). The region that is initially bonded is indicated by the red rectangle. (b) Normalized shear rate $\dot{\gamma}_{MSDWS}/\dot{\gamma}_{macro}$ calculated for the shaded region in (a), as a function of the distance y from the top edge. Curves are offset vertically for clarity, the bar

*corresponds to a change of 0.3 in $\dot{\gamma}_{MSDWS}/\dot{\gamma}_{macro}$. Data are color-coded according to the value of $\dot{\gamma}_{MSDWS}/\dot{\gamma}_{macro}$, as shown by the color bar on the right.*

## B. Dynamic activity maps

The results shown in Fig. 8 were obtained by calculating the local shear rate from the characteristic time $\tau_0$ determined by a fit of the full autocorrelation function. This procedure leads to excellent results, but is somehow time consuming. Indeed, for each $t$ and each position $\vec{r}$ of interest one needs to first calculate $C(t, \tau, \vec{r})$ for a sufficiently large sets of time delays, and then perform a non-linear fit on each $C(\tau)$ function. To speed up the processing, for each $t$ one may simply calculate a dynamic activity map (DAM), i.e. a spatial map of the degree of correlation at a fixed time delay $\tau$ [31]. Assuming that the decay of the autocorrelation function may be attributed only to the affine deformation and taking for simplicity $a = 0$, from Eqs. 3, 4 and 10 one has

$$\left(-\frac{\ln\frac{C(\tau)-C(\infty)}{C(0)-C(\infty)}}{2\epsilon}\right)^2 = 3k^2 l^{*2} f[\boldsymbol{D}(\tau)] = \frac{\tau^2 \dot{\gamma}_{MSDWS}^2 k^2 l^{*2}}{10}$$

$$C(\infty) = B, C(0) - C(\infty) = A, \tag{17}$$

such that

$$\dot{\gamma}_{MSDWS} = \frac{\sqrt{10}\ln\frac{C(0)-C(\infty)}{C(\tau)-C(\infty)}}{2\epsilon k l^* \tau}. \tag{18}$$

In Eq. 18, $C(0)$ and $C(\infty)$ are the correlation values at zero delay and infinite delay, respectively, which can be approximately calculated as the correlation of one image with itself and with images at a large time delay $\tau$. In writing Eq. 18, we have used the expression of $\boldsymbol{D}$ for simple shear: one can write similar equations for different deformation geometries, e.g. for uniaxial extension using Eq. 9. Using Eq. 18, the DAMs can thus be directly converted to maps of the local strain rate[84, 85]. This procedure is much faster than the fitting scheme described above, essentially because the correlation coefficients $C$ need to be calculated only for one lag value.

Fig 9(a) shows DAMs calculated at $t = 1080$ s for $\tau = 0.18$ s and $0.6$ s, again for the experiment with $\sigma_{macro}$ = 0.62 MPa. The two maps show qualitatively similar dynamics distributions: at the top-right corner where the debonding starts, the value of $C(\tau)$ is lower, implying that the dynamics are faster. Note that, by choosing different values of the time delay $\tau$, one can tune the sensitivity of the DAM to deformations occurring on different time scales, striking the desired balance between measuring the "instantaneous" deformation (by keeping $\tau$ as small as possible) and assessing the cumulated deformation (by increasing $\tau$). Fig 9(b), compares the maps of the local strain obtained either from the DAM shown in Fig 9(a), or by fitting the full (local) autocorrelation functions. The spatial distributions obtained from the two methods are very similar, demonstrating the soundness of the approach based on the DAMs. For these data, the gain in time when using the DAMs is quite substantial, since the full fitting procedure was about 100 times slower than that based on the DAMs.

A few remarks on the limits of the DAM approach are of interest. First, we note that retrieving the deformation rate from a DAM is only valid when the autocorrelation functions have the same shape over the whole image. Furthermore, when the magnitude of the deformation rate varies too much over the imaged sample or with time, the DAM approach significantly loses accuracy, due to the exponential shape of the autocorrelation function, which makes the DAMs poorly sensitive to deformation for time delays too small or too large compared to the relaxation time of $C$. Finally, the amplitude of the autocorrelation function, $A$ in Eq. 3, depends on the coherence of the scattered light, which may vary over the imaging field. In practice, the most reliable and efficient approach would mix both fitting and using DAMs: (1) correlation functions calculated in real time on a subset of the acquired images should be monitored, in order to determine the optimal $\tau$ for DAMs. (2) Renormalization of the correlation value by spatially resolved $A$ in the subset images.

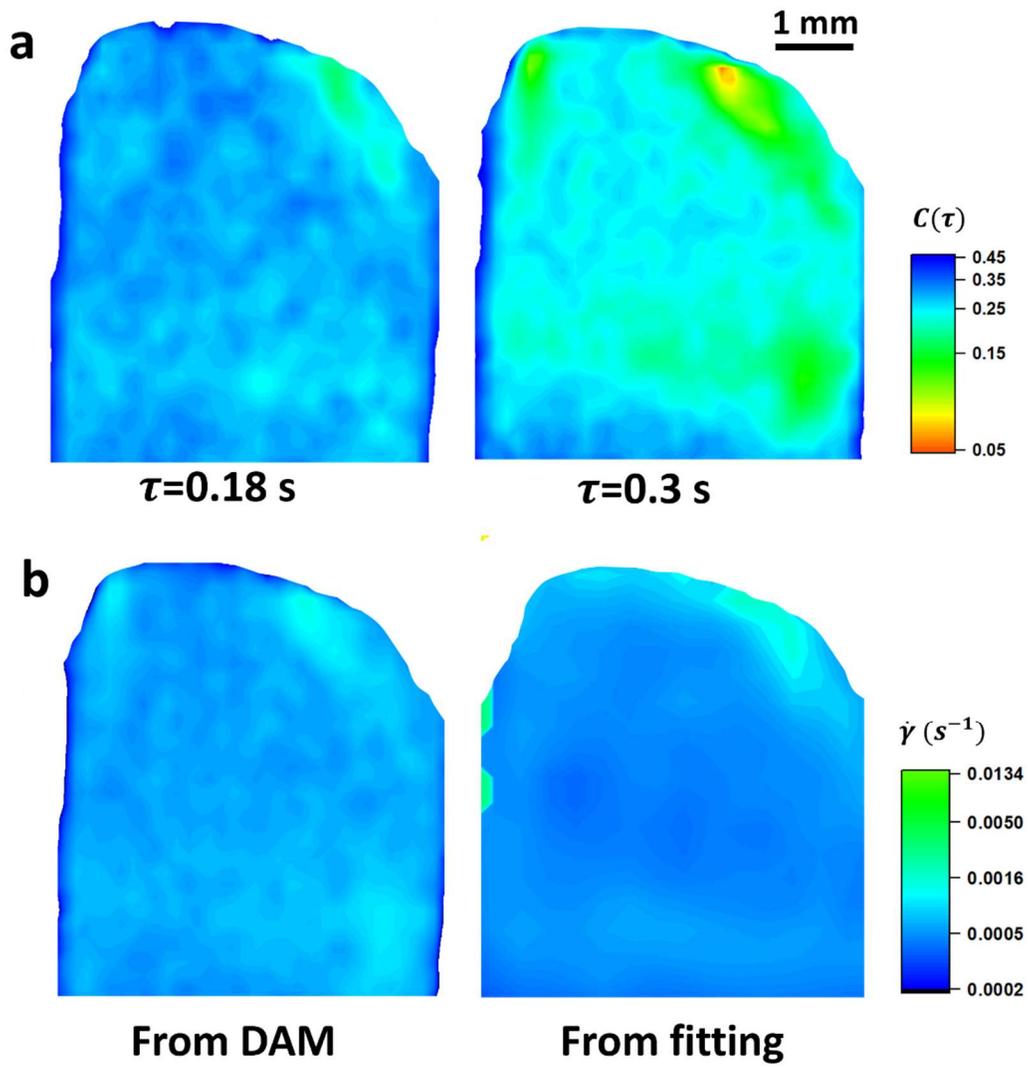

*Fig. 9 (a) Dynamic activity maps (DAMs) for $\tau = 0.18$ s (left) and 0.3 s (right). (b) $\dot{\gamma}_{MSDWS}$ map calculated from a DAM (left) and from fits of the local autocorrelation functions.*

## VII. Conclusions

We have shown that MSDWS is a powerful tool for the measurement of spatially- and temporally-varying mechanical responses. Even though a complete modelling of the intensity autocorrelation function remains challenging, we have demonstrated that in a wide range of common testing situations it is possible to directly quantify the deformation rate distribution from the decay of space- and time-resolved correlation functions. MSDWS is relatively simple, requiring only illumination of the sample with coherent laser light and detection with a 2D sensor, typically a standard CMOS camera. Because MSDWS is not too sensitive to the exact illumination and detection parameters (angles of the incident and collected light with respect to the sample, acceptance angular range of the detector etc.), the requirements on optical alignment and the experimental environment are less strict than in other optical methods. The tests on PDMS under elongation strain have allowed us to validate the theoretical analysis presented in section II. The method was then applied to the investigation of adhesive debonding, unveiling a heterogeneous deformation rate distribution and the emergence of microscopic precursors of failure hundreds of seconds before macroscopic debonding.

This paper provides theoretical and practical guidelines for using MSDWS as a powerful tool for measuring the local mechanical response of a loaded soft solid. A successful and efficient implementation requires a suitable combination of sample opacity ($l^*$), range of deformation rate, testing geometry and data processing strategy. We hope that our work will further spur interest in this powerful method.

## Acknowledgements

This work was supported by the European Union's Horizon 2020 Programme for Research and Innovation under the Marie Skłodowska-Curie grant agreement no. 765811 (DoDyNet) and by the French ANR (ANR-20-CE06-0028, MultiNet). L.C. acknowledges support from the Institut Universitaire de France. We thank Chung Yuen Hui for valuable discussions and M.-Y. Nagazi for preliminary tests on PDMS.

## Data Availability

ASCII and tif. format images shown in the figures in the main text and SI have been uploaded in Zenodo (DOI: 10.5281/zenodo.5912067).

# Supporting Information

# Multispeckle diffusing wave spectroscopy as a tool to study heterogeneous mechanical behavior in soft solids

Jianzhu Ju [a], Luca Cipelletti [b,c], Stephan Zoellner [d], Tetsuharu Narita*[,a,e], Costantino Creton*[,a,e]

[a] Sciences et Ingénierie de la Matière Molle, CNRS UMR 7615, ESPCI Paris, PSL Université, Paris, France

[b] Laboratoire Charles Coulomb (L2C), University of Montpellier, CNRS, Montpellier, France

[c] Institut Universitaire de France

[d] TESA SE, 22848 Norderstedt, Germany

[e] Global Station for Soft Matter, Global Institution for Collaborative Research and Education, Hokkaido University, Sapporo, Japan

## 1. Details of the parameters in the section II B (Eq. 2)

$$C(\tau)_{\vec{r}_0, t_0} = A\exp\left(-2\epsilon\sqrt{\left(\frac{\tau}{\tau_0}\right)^p + a}\right) + B. \qquad (2)$$

- $\epsilon$ is related to the polarization of the incident and detected light. We used a typical value being $\epsilon = 1.8$ when no polarizer is used before detection [20].

- $a = 3l^*/l_a$ depends on the ratio of the photon transport mean free path $l^*$ and the absorption length $l_a$ (usually, $a \ll 1$), which needs to be experimentally determined [20, 34]. When the sample absorb the light at the wavelength of the laser, the absorption needs to be considered. Also the scattered light intensity decreases.

- $\tau_0$ is the characteristic decay time, corresponding to the time scale over which $C$ decreases to around $exp(-2\epsilon)$ after normalization, and the scatterers are displaced over a distance of the order of $1/k$. (typically, around 50 nm). $k$ is the wave vector of the laser light defined as $k = 2n\pi/\lambda$, with $n$ the refractive index of the material and $\lambda$ the in-vacuum laser wavelength.

*correspondence authors: tetsuharu.narita@espci.fr, costantino.creton@espci.fr

The value of $1/k$ is typically around 50 nm.

• $B$ is the baseline of the correlation function: ideally, $B=0$, but measurement noise and other artifacts often result in a non-zero baseline [35]. In practice, $B$ is determined from the long-time behavior of $C$ and subtracted from the experimental curve. $A$ is a normalization factor set by imposing $C(\tau)_{\vec{r}_0,t_0} - B \to 1$ for $\tau \to 0$.

## 2. Uniaxial tensile stage

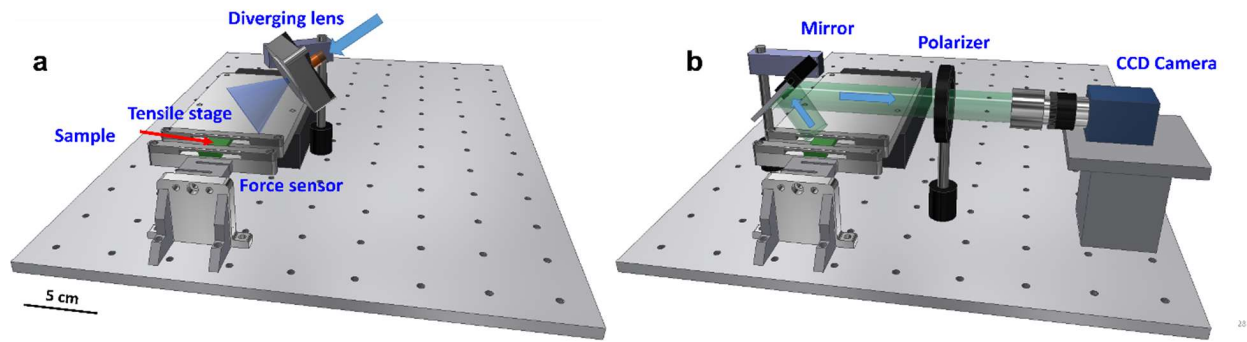

*Fig. S1 Detailed image of the uniaxial tensile stage (a) Illumination part. (b) Scattered light collection part.*

## 3. Fitting of autocorrelation function

An example of fitting of the autocorrelation function during the uniaxial extension of dogbone PDMS sample ($\lambda_N = 1.16$) is shown in Fig. S2. $a$ is set to be constant and equal to 0.015 for all data, as absorption and finite thickness corrections may be regarded as being essentially independent of $\lambda_N$.

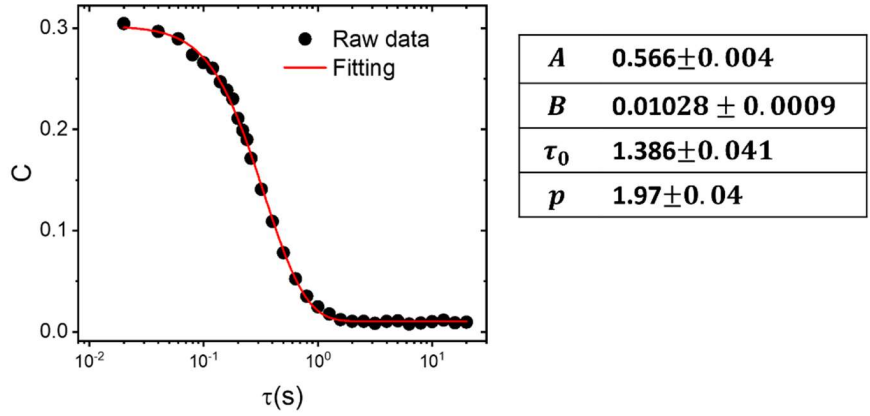

Fig. S2 Example of fitting of the autocorrelation function measured for an uniaxially stretched PDMS elastomer, at $\lambda_N = 1.16$.

## 4. Fitting of $l^*$ in adhesive tapes

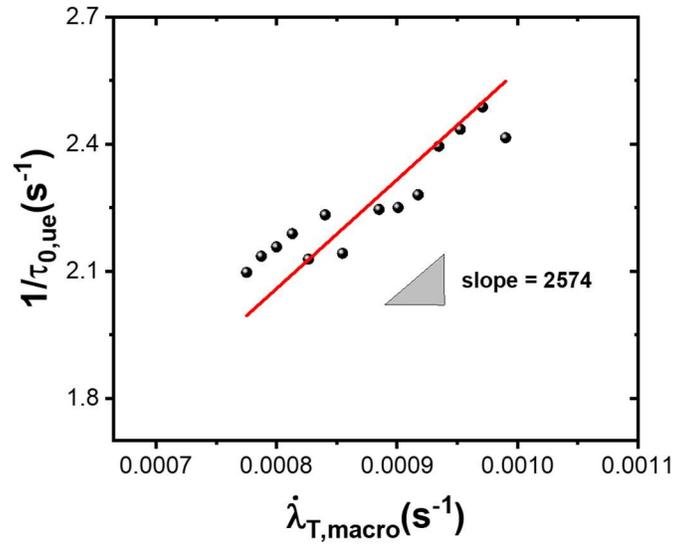

Fig. S3 $1/\tau_{0,ue}$ as a function of $\dot{\lambda}_{T,macro}$, measured in tapes under uniaxial tension.

$\tau_{0,ue}$ of the adhesive tape is measured under uniaxial tension with the set-up shown in Fig. 2(a) in the main text, with $\dot{\lambda}_N = 0.001 \, s^{-1}$. To obtain $l^*$ from uniaxial tensile testing, $1/\tau_{0,ue}$ is plotted as a function of $\dot{\lambda}_{T,macro}$ in Fig. S3 calculated using Eq. 9 of the main text. $1/\tau_{0,ue}$

should be proportional to $\dot{\lambda}_{T,macro}$, through a proportionality coefficient that scales as $kl^*$ (see Eq. 9). From Fig. S3, a slope of 2574 is obtained, corresponding to $kl^* = 3.3\times10^3$.